\documentclass[aps,prl,twocolumn,showpacs,showkeys,footinbib,superscriptaddress]{revtex4-1}

\usepackage{amsmath}
\usepackage{amssymb}
\usepackage{graphicx}
\usepackage{bm}
\usepackage{color}
\usepackage{relsize}
\usepackage{braket}
\usepackage[caption=false]{subfig}

\usepackage{hyperref}
\usepackage[all]{hypcap}

\renewcommand{\i}{\ensuremath{\mathrm{i}}}

\renewcommand{\d}{\ensuremath{\mathrm{d}}}

\begin{document}
\title{Magnetic Proximity Effects in Transition-Metal Dichalcogenides: Converting Excitons}
\date{\today}

\author{Benedikt Scharf}
\affiliation{Institute for Theoretical Physics and Astrophysics, University of W\"{u}rzburg, Am Hubland, 97074 W\"{u}rzburg, Germany}
\affiliation{Institute for Theoretical Physics, University of Regensburg, 93040 Regensburg, Germany}
\affiliation{Department of Physics, University at Buffalo, State University of New York, Buffalo, New York 14260, USA}
\author{Gaofeng Xu}
\affiliation{Department of Physics, University at Buffalo, State University of New York, Buffalo, New York 14260, USA}
\author{Alex Matos-Abiague}
\affiliation{Department of Physics, University at Buffalo, State University of New York, Buffalo, New York 14260, USA}
\author{Igor \v{Z}uti\'c}
\affiliation{Department of Physics, University at Buffalo, State University of New York, Buffalo, New York 14260, USA}
\begin{abstract} 
The two-dimensional character and reduced screening in monolayer transition-metal dichalcogenides (TMDs) lead to the ubiquitous formation of robust excitons with binding energies orders of magnitude larger than in bulk semiconductors. Focusing on neutral excitons, bound electron-hole pairs, that dominate the optical response in TMDs, it is shown that they can provide fingerprints for magnetic proximity effects in magnetic heterostructures. These proximity effects cannot be described by the widely used single-particle description, but instead reveal the possibility of a conversion between optically inactive and active excitons by rotating the magnetization of the magnetic substrate. With recent breakthroughs in fabricating Mo- and W-based magnetic TMD heterostructures, this emergent optical response can be directly tested experimentally.
\end{abstract}


\maketitle
Proximity effects can transform a given material through its adjacent regions to become superconducting, magnetic, or topologically nontrivial~\cite{DeGennes1964:RMP,Buzdin2005:RMP,Wang2015:PRL,Fu2008:PRL,Gmitra2016:PRB,Efetov2015:NP,Hauser1969:PR,Lazic2014:PRB,Lazic2016:PRB}. In bulk materials, the sample size often dwarfs the characteristic lengths of proximity effects allowing their neglect. However, in monolayer (ML) van der Waals materials such as graphene or transition-metal dichalcogenides (TMDs), the situation is drastically different, even short-range magnetic proximity effects exceed their thickness~\cite{Hauser1969:PR,Lazic2014:PRB}.

MX$_2$ (M = Mo,W, X = S, Se, Te) ML TMDs have unique optical properties that combine a direct band gap, very large excitonic binding energies (up to $\sim$0.5 eV), and efficient light emission~\cite{Mak2010:PRL,Chernikov2014:PRL,Amani2015:S}. A hallmark of TMDs is their valley-spin coupling which leads to a valley-dependent helicity of interband optical transitions as well as important implications for transport and qubits~\cite{Xu2014:NP,Mak2014:S,Wang2012:NN,Dery2016:PRB,Klinovaja2013:PRB,Rohling2014:PRL}. Lifting the degeneracy between the valleys $K$ and $K'$ was identified as the key step in manipulating valley degrees of freedom. However, a common approach was focused on very large magnetic fields required by a small Zeeman splitting of $\sim$0.1-0.2 meV/T~\cite{Srivastava2015:NP,MacNeill2015:PRL,Stier2016:NC,Plechinger2016:NL,Arora2016:NL}. Instead, recent experimental breakthroughs demonstrate a viable alternative by using optically detected magnetic proximity effects dominated by excitons~\cite{Zhao2017:NN,Zhong2017:SA,Ye2016:NN}.

While a single-particle description already suggests unusual implications of magnetic proximity effects~\cite{Lazic2016:PRB,Qi2015:PRB}, strong many-body interactions 
qualitatively alter the optical response in TMDs and yield a wealth of unexplored phenomena~\cite{Xu2014:NP,vanTuan2017:P}. Here, we provide the missing description of Coulomb interaction in magnetic proximity effects and elucidate how they transform the observed excitons in TMDs on magnetic substrates. In the seemingly trivial case of an in-plane magnetization, $\bm{M}$, where a single-particle description implies no lifting of the valley degeneracy~\cite{Qi2015:PRB}, we predict that dark neutral excitons $X^0$ become bright. The term dark (bright) represents optically forbidden (allowed) dipole transitions with an antiparallel (parallel) electron spin configuration.

\begin{figure}[ht]
\centering
\includegraphics*[width=8.6cm]{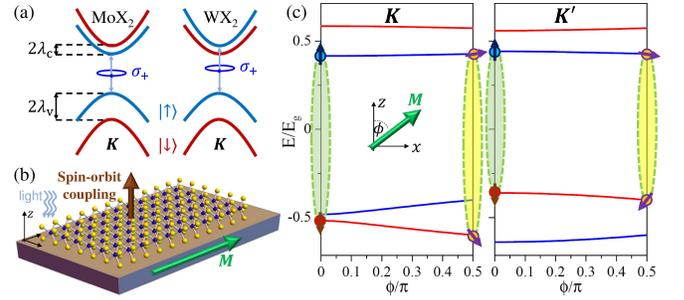}
\vspace{-0.5cm}
\caption{(a) Spin-valley coupling in ML TMDs: The conduction (valence) band is spin-split in the $K$ valley by $2\lambda_\mathrm{c(v)}$ due to strong spin-orbit coupling with the CB ordering reversed between MoX$_2$ and WX$_2$; at the $K'$ point all spin orientations are reversed (not shown); emitted/absorbed light has valley-selective helicity, $\sigma_\pm$. (b) ML TMD on a magnetic substrate. (c) The $K$ and $K'$ band edges as the substrate's magnetization, $\bm{M}$, is rotated, shown for MoTe$_2$/EuO parameters. One dark exciton for $K$ and $K'$ and the spin direction for selected band edges are depicted.}
\label{fig:Scheme}
\vspace{-0.3cm}
\end{figure}

Figure~\ref{fig:Scheme}(a) shows the band structure of ML TMDs reflecting strong spin-orbit coupling (SOC) due to the $d$ orbitals of the heavy metal atoms and broken inversion symmetry and the considered geometry. For bands with a 2D representation, the SOC Hamiltonian can be written as $H_\mathrm{SO}=\bm{\Omega}(\bm{k})\cdot\bm{s}$ using the SOC field $\bm{\Omega}(\bm{k})$~\cite{Zutic2004:RMP,Fabian2007:APS}, where $\bm{k}$ is the wave vector and $\bm{s}$ is the vector of spin Pauli matrices. In ML TMDs, this leads to $\bm{\Omega}(\bm{k})=\lambda(\bm{k})\hat{\bm{z}}$, where $\lambda(\bm{k})$ is odd in $\bm{k}$ and $\hat{\bm{z}}$ is the unit vector normal to the ML plane. At the $K$ point, $\lambda(\bm{k})$ reduces to the values $\lambda_\mathrm{c(v)}$ in the conduction (valence) band CB (VB), Fig.~\ref{fig:Scheme}(a). The limiting case of this description with a magnetic substrate, see Fig.~\ref{fig:Scheme}(b), neglecting many-body effects is given by the Hamiltonian~\cite{Xiao2012:PRL,Qi2015:PRB} ${H}_\mathrm{tot}={H}_0+{H}_\mathrm{ex}+{H}_\mathrm{R}$, a sum of the ``bare" ML TMD, a proximity-induced exchange term, and Rashba SOC with
\begin{eqnarray}
H_\mathrm{0} &=& \hbar v_\mathrm{F}(k_x\sigma_x\tau_z+k_y\sigma_y)+(E_\mathrm{g}/2)\sigma_z\\ 
\nonumber
&+& {\tau}_z\ {s}_z\left[\lambda_\mathrm{c}(1+{\sigma}_z)/2+\lambda_\mathrm{v}(1-{\sigma}_z)/2\right],
\label{eq:H0}
\end{eqnarray}
where $\sigma_i$ and ${\tau}_i$ denote Pauli matrices for the CB/VB and the valley, $v_\mathrm{F}$ the Fermi velocity, and $E_\mathrm{g}$ the band gap in the absence of SOC. Writing $\bm{M}=M\hat{\bf n}$, we have
\begin{equation}
H_\mathrm{ex}=-\hat{\bf n}\cdot{\bf s}\left[J_{\mathrm{c}}(1+{\sigma}_z)/2+J_{\mathrm{v}}(1-{\sigma}_z)/2\right],
\label{eq:Hex}
\end{equation}
where $J_{\mathrm{c}(\mathrm{v})}$ is the exchange splitting induced in the CB (VB), while in the Rashba SOC, ${H}_\mathrm{R}=\lambda_\mathrm{R}({s}_y{\sigma}_x{\tau}_z-{s}_x{\sigma}_y)$, $\lambda_\mathrm{R}$ is the Rashba SOC parameter. 

From the resulting single-particle description, $H_\mathrm{tot}\eta^{\tau}_{n\bm{k}}=\epsilon^\tau_n(\bm{k})\eta^{\tau}_{n\bm{k}}$ with the energies $\epsilon^\tau_n(\bm{k})$ and the corresponding eigenstates $\eta^{\tau}_{n\bm{k}}$, we develop a generalized Bethe-Salpeter equation (BSE) to elucidate many-body manifestations of magnetic proximity effects. The BSE can be conveniently written as~\cite{Rohlfing2000:PRB,Scharf2016:PRB,Note:SM}
\begin{equation}
[\Omega^{\tau}_{S} -\epsilon^\tau_c(\bm{k})+\epsilon^\tau_v(\bm{k})] \mathcal{A}^{S\tau}_{vc\bm{k}} =\sum\limits_{v'c'\bm{k}'}\mathcal{K}^{\tau}_{vc\bm{k},v'c'\bm{k}'}\mathcal{A}^{S\tau}_{v'c'\bm{k}'},
\label{eq:BSE}
\end{equation}
where in a given valley $\tau$ the band index $n=c(v)$ denotes one of the two CBs (VBs), $\Omega^{\tau}_{S}$ is the energy of the exciton state $\ket{\Psi^{\tau}_{S}}=\sum_{vc\bm{k}}\mathcal{A}^{S\tau}_{vc\bm{k}}\hat{c}^\dagger_{\tau c\bm{k}}\hat{c}_{\tau v\bm{k}}\ket{\mathrm{GS}}$~\cite{footnote:BSE} with the coefficients $\mathcal{A}^{S\tau}_{vc\bm{k}}$, the creation (annihilation) operator of an electron in a CB $c$ (VB $v$) $\hat{c}^\dagger_{\tau c\bm{k}}$ ($\hat{c}_{\tau v\bm{k}}$) in this valley, and the ground state $\ket{\mathrm{GS}}$ with fully occupied VBs and unoccupied CBs. Here, the kernel, $\mathcal{K}^{\tau}_{vc\bm{k},v'c'\bm{k}'}$, includes the Coulomb interaction between electrons in the layer, determined from the dielectric environment, geometry, and form factors calculated from $\eta^{\tau}_{n\bm{k}}$~\cite{Keldysh1979:JETP,Cudazzo2011:PRB,Note:SM}. The influence of magnetic substrates therefore modifies not only the single-particle energies $\epsilon^\tau_{c/v}(\bm{k})$, but also the many-body interactions through this $\bm{M}$-dependent kernel~\cite{Note:SM}, which could be generalized to include other quasiparticle excitations beyond excitons.

While experiments demonstrate the proximity-induced exchange splitting in ML TMDs using an adjacent ferromagnet~\cite{Zhao2017:NN,Zhong2017:SA,Ye2016:NN}, the employed single-particle description poses large uncertainties and excludes detected excitons. Equation~(\ref{eq:BSE}) now allows us to compute excitons in TMDs as $\bm{M}$ is rotated. Generally, $\mathcal{K}^{\tau}_{vc\bm{k},v'c'\bm{k}'}$ couples all CBs and VBs in a valley. Only if spin is a good quantum number, in the absence of Rashba SOC and $\bm{M}\|\hat{\bm{z}}$, do the CBs (VBs) decouple. Each exciton is then formed from only one CB and VB and can be labeled by the spin configuration of those bands. This is no longer exactly true for arbitrary $\bm{M}$ orientation, but our results~\cite{Note:SM} show that typically the coupling between different CBs (VBs) is small and excitons are still mainly formed from one specific CB and VB as depicted in Fig.~\ref{fig:Scheme}(c).

\begin{figure}[ht]
\centering
\includegraphics*[width=8.6cm]{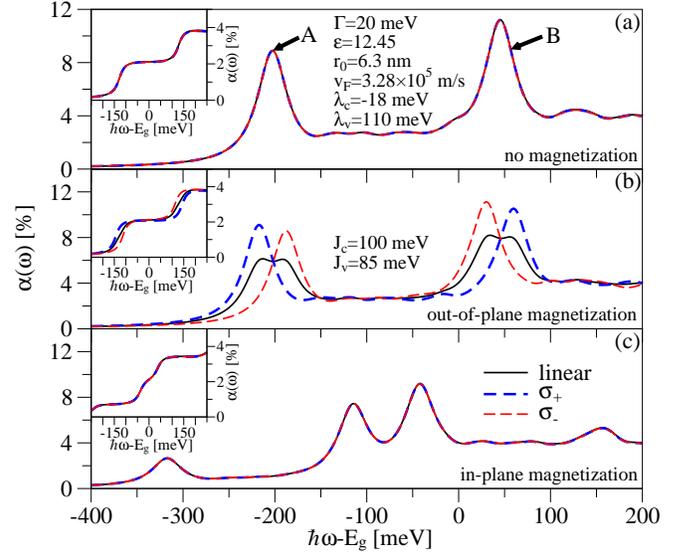}
\vspace{-0.5cm}
\caption{Absorption spectra of ML MoTe$_2$ on EuO for different polarizations if (a) no proximity-induced exchange splitting, (b) out-of-plane and (c) in-plane exchange splitting are considered. The insets show the respective single-particle absorptions computed without excitonic effects. $E_\mathrm{g}=1.4$ eV.}\label{fig:Comparison}
\vspace{-0.3cm}
\end{figure}

The proximity-modified optical response, including excitons, can be accurately studied through the absorption,
\begin{equation}
\alpha(\omega)=\frac{4e^2\pi^2}{c\omega}\frac{1}{A}\sum\limits_{S\tau}\left|\sum\limits_{vc\bm{k}}\mathcal{D}_{vc\bm{k}}\mathcal{A}^{S\tau}_{vc\bm{k}}\right|^2\delta(\hbar\omega-\Omega^{\tau}_{S}),
\label{eq:absorption}
\end{equation}
where $\omega$ is the photon frequency of light propagating along the $-\hat{\bm{z}}$ direction, $c$ the speed of light, $A$ the 2D unit area, and the single-particle velocity matrix elements are given by $\mathcal{D}^{\sigma^\pm}_{vc\bm{k}}=\left[\eta^{\tau}_{c\bm{k}}\right]^\dagger\hat{v}_\pm\eta^{\tau}_{v\bm{k}}$ for circularly polarized light and by $\mathcal{D}^{x/y}_{vc\bm{k}}=\left[\eta^{\tau}_{c\bm{k}}\right]^\dagger\hat{v}_{x/y}\eta^{\tau}_{v\bm{k}}$ for linearly polarized light with $\hat{v}_\pm=(\hat{v}_x\pm\i\hat{v}_y)/\sqrt{2}$, $\hat{v}_{x/y}=\partial H_\mathrm{tot}/\partial(\hbar k_{x/y})$. Optically inactive excitons or excitations imply that $\sum\limits_{vc\bm{k}}\mathcal{D}_{vc\bm{k}}\mathcal{A}^{S\tau}_{vc\bm{k}}$ has to vanish---due to either orbital restrictions or the spin configuration. The $\delta$ function is modeled by a Lorentzian with broadening $\Gamma$.

A common approach for robust magnetic proximity effects in 2D materials is to minimize the hybridization effects and employ a magnetic insulator or a semiconductor~\cite{Zhao2017:NN,Zhong2017:SA,Ye2016:NN,Wei2016:NM,Swartz2012:ACSN,Yang2013:PRL}. First-principles results suggest a giant proximity-induced exchange splitting in MoTe$_2$/EuO~\cite{Qi2015:PRB}, which has also guided our choice of parameters. We use a reduced exchange coupling of $J_\mathrm{c}=100$ meV and $J_\mathrm{v}=85$ meV to reflect the fact that the calculated (111) interface is polar~\cite{Qi2015:PRB} and will undergo interface reconstruction~\cite{Note:Kirill}. The use of the optical response can provide a cleaner detection of magnetic proximity effects than through transport measurements, which can be complicated by various artifacts and complex interfaces. Similar transport difficulties are already known from the case of spin injection and detection~\cite{Zutic2004:RMP,Song2014:PRL}. Using ML MoTe$_2$ parameters~\cite{Kormanyos2015:2DM,footnote:MoTe2Parameters}, a background dielectric constant $\varepsilon=12.45$ and the ML polarizability parameter $r_0=6.3$ nm to model ML MoTe$_2$ on EuO~\cite{Note:SM}, we set $\lambda_\mathrm{R}=0$ in the following for simplicity. However, we find that Rashba SOC does not significantly change our results~\cite{Note:SM}.

Similar to many experiments on ML TMDs~\cite{Mak2010:PRL,Xu2014:NP}, the calculated absorption for $\bm{M}=\bm{0}$ in Fig.~\ref{fig:Comparison}(a) is polarization independent and dominated by the so-called A and B peaks of bright neutral excitons $X^0$, corresponding to dipole-allowed transitions from the upper (A) and lower (B) valence band, respectively [see also Fig.~\ref{fig:Scheme}(a)]. In contrast, these peaks are completely absent in the single-particle picture [insets in Figs.~\ref{fig:Comparison}(a)-(c)], which is insufficient to properly include the excitonic effects~\cite{YuCardona2010}.

The polarization independence of the absorption is a consequence of the valley degeneracy, lifted by an out-of-plane $\bm{M}$ as seen in Fig.~\ref{fig:Comparison}(b). To understand the removal of the valley degeneracy, we focus on circularly polarized light because $\sigma_+$ ($\sigma_-$) couples exclusively to the $K$ ($K'$) valley. Since the exchange splitting is different for the CBs and VBs (with $J_\mathrm{c}>J_\mathrm{v}>0$), the single-particle gap energy for spin-up (spin-down) transitions at $K$ is decreased (increased), resulting in a redshift of the A peak and a blueshift of the B peak for $\sigma_+$. An opposite behavior for $\sigma_-$ is seen at $K'$, leading to a splitting between $\sigma_+$ and $\sigma_-$ of 29 meV and 30 meV for the A and B peaks, respectively. The absorption for $x$-polarized light is the symmetric combination of $\sigma_+$ and $\sigma_-$.

Despite the common valley degeneracy and polarization independence in Figs.~\ref{fig:Comparison}(a) and (c), there is a striking difference in the position of the
two main peaks and the emergence of a new low-energy peak for $\bm{M}\perp\hat{\bm{z}}$. A clue for this behavior comes from Fig.~\ref{fig:Scheme}(c). While there are well-defined dark and bright excitons for an out-of-plane $\bm{M}$, the situation changes when $\bm{M}$ is rotated in plane. As the spin projections of a CB and VB forming a dark exciton at $\bm{M}\|\hat{\bm{z}}$ are no longer perfectly antiparallel if $\bm{M}$ acquires an in-plane component, the single-particle matrix elements $\mathcal{D}_{vc\bm{k}}$ between these two bands and hence also the exciton dipole matrix element become finite and the formerly dark excitons become bright. 

\begin{figure}[ht]
\centering
\includegraphics*[width=8.6cm]{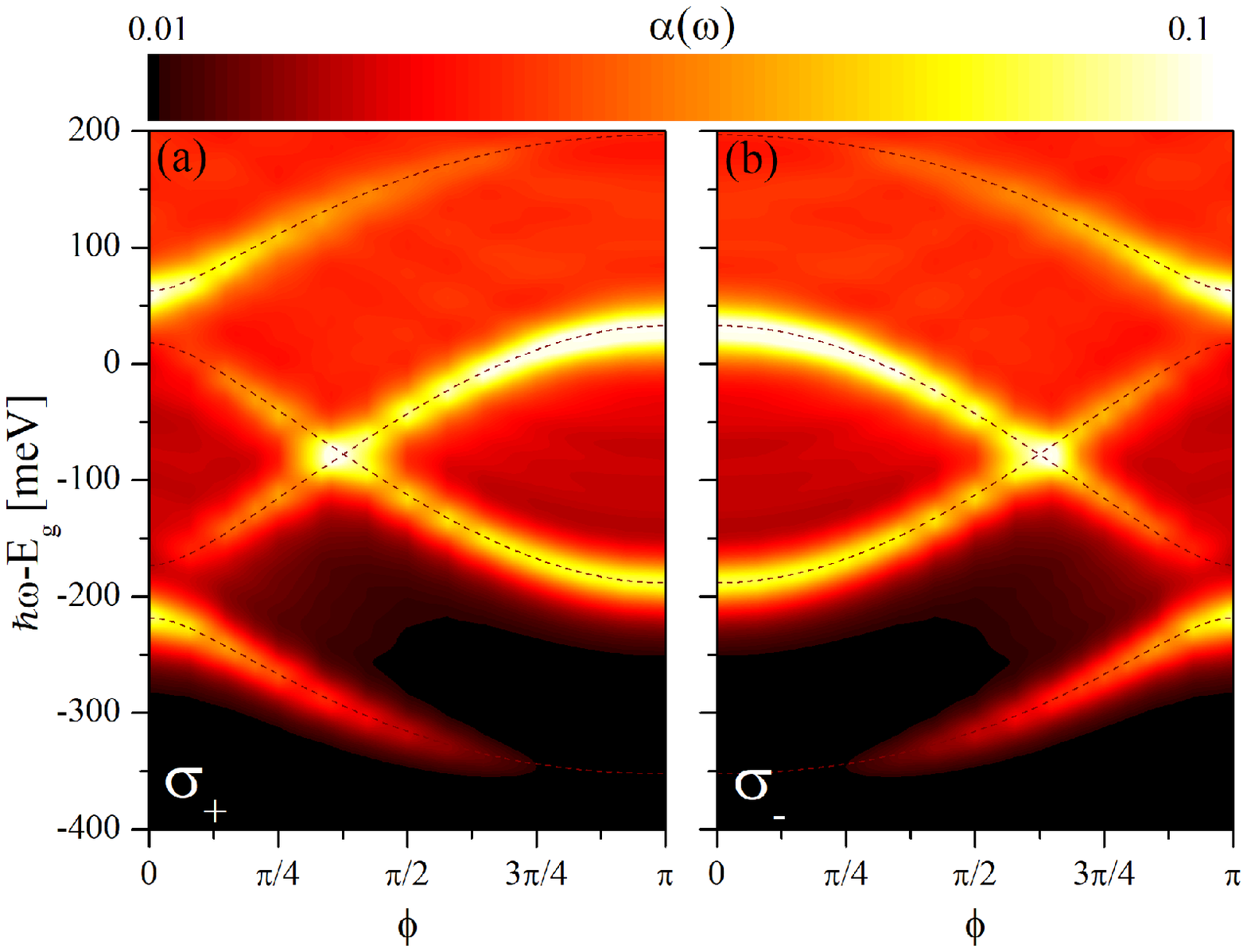}
\vspace{-0.5cm}
\caption{Evolution of the absorption of ML MoTe$_2$ on EuO as ${\bm M}$ is rotated from out of plane ($\phi=0$) to in plane ($\phi=\pi/2$) and out of plane, but with reversed ${\bm M}$ ($\phi=\pi$) for (a) $\sigma_+$ and (b) $\sigma_-$. The dashed lines are the peak positions from Eq.~(\ref{eq:peakposition}). The parameters are from Figs.~\ref{fig:Comparison}(b) and~(c).}
\label{fig:Rotation}
\vspace{-0.3cm}
\end{figure}

Following the rotation of ${\bm M}$ from $\phi=0$ to $\pi$ (recall Fig.~\ref{fig:Scheme}) illustrates in Fig.~\ref{fig:Rotation} a peculiar transfer of spectral weight. In addition to the two bright excitons at $\phi=0$, with an increase in $\phi$ two dark excitons gradually become bright, consistent with four bright excitons at $\phi=\pi/2$ in Fig.~\ref{fig:Comparison}(c). Such a brightening of excitons due to spin flips, also predicted due to out-of-plane electric fields~\cite{Dery2015:PRB}, has recently been observed in very large in-plane magnetic fields, $B \approx 30$ T~\cite{Molas2017:2DM,Zhang2017:NN}, while found negligible even at $B>8$ T~\cite{Srivastava2015:NP} in ML TMDs. However, Fig.~\ref{fig:Rotation} shows also a reverse process: As $\phi$ is increased, there is a darkening of the bright excitons. By reversing ${\bm M}$, there is a complete conversion between the dark and bright excitons.

Together with the spectral weight transfer, Fig.~\ref{fig:Rotation} reveals that all exciton peaks are shifted in energy. From the full BSE calculation, it is possible to obtain a simplified description based on the $\phi$ evolution of the single-particle optical gap, neglecting the ${\bm M}$-dependent changes in the exciton binding energy, which in other cases can be significant as shown in Fig.~S4 in Ref.~\onlinecite{Note:SM}. The corresponding peak positions, marked by dashed lines in Fig.~\ref{fig:Rotation}, are given by
\begin{equation}
\hbar\omega^\mathrm{exc}_{cv}(\phi)=\epsilon^\tau_c(\phi)-\epsilon^\tau_v(\phi)-E^{cv}_\mathrm{b},
\label{eq:peakposition}
\end{equation}
where $E^{cv}_\mathrm{b}$ is the binding energy of the $1s$ exciton formed from CB $c$ and VB $v$ at $\phi=0$, which does not significantly change for $\phi\neq0$ here, and $\epsilon^\tau_c(\phi)$ and $\epsilon^\tau_v(\phi)$ are their respective single-particle band edges at $\bm{k}=\bm{0}$. For $\lambda_\mathrm{R}=0$, these can be computed analytically as
\vspace{-0.1truecm}
\begin{eqnarray}
\epsilon^\tau_{\mathrm{v},\pm}(\phi)=-(E_\mathrm{g}/2)\pm\sqrt{J^2_\mathrm{v}\sin^2\phi+\left(\tau\lambda_\mathrm{v}-J_\mathrm{v}\cos\phi\right)^2},\nonumber\\
\epsilon^\tau_{\mathrm{c},\pm}(\phi)=(E_\mathrm{g}/2)\pm\sqrt{J^2_\mathrm{c}\sin^2\phi+\left(\tau\lambda_\mathrm{c}-J_\mathrm{c}\cos\phi\right)^2}\quad\,
\label{eq:SPbandedges}
\vspace{-0.1truecm}
\end{eqnarray}
for the two VBs $v=(\mathrm{v},\pm)$ and the two CBs $c=(\mathrm{c},\pm)$. The peak positions calculated from Eq.~(\ref{eq:peakposition}) agree well with those computed from the full BSE in Eq.~(\ref{eq:BSE}), demonstrating the usefulness of this simplified picture.

Symmetry requires the energy spectra (and consequently the absorption) at the $K$ and $K'$ points to be connected via $\epsilon^{-\tau}_n(\phi)=\epsilon^{\tau}_n(\pi-\phi)$, which is also reflected by Eq.~(\ref{eq:SPbandedges}) and Figs.~\ref{fig:Rotation}(a) and (b). With our parameters for MoTe$_2$ on EuO, the strong exchange splitting in the CB determines the spin ordering in both valleys, while the VB spin ordering is unaffected [see Fig.~\ref{fig:Scheme}(c)]. Hence, the exciton with lowest energy in the $K$ valley is bright for $\phi=0$, whereas it is dark for $\phi=\pi$. Conversely, the exciton with lowest energy in the $K'$ valley is dark for $\phi=0$ and bright for $\phi=\pi$. 

\begin{figure}[ht]
\centering
\includegraphics*[width=8.6cm]{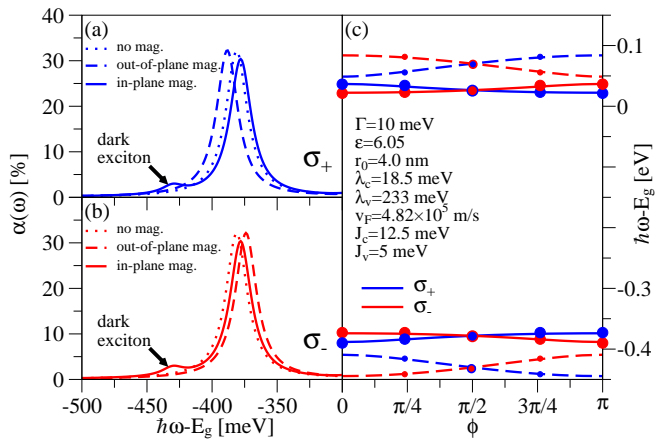}
\vspace{-0.5cm}
\caption{Absorption profile of the A peak in ML WSe$_2$ on EuS for (a) $\sigma_+$ and (b) $\sigma_-$ and different ${\bm M}$ directions. (c) The evolution of the bright (dark) excitons as solid (dashed) lines as ${\bm M}$ is reversed from $\phi=0$ to $\phi=\pi$. The size of the circles represents exciton oscillator strengths for selected orientations of ${\bm M}$ (not to scale).}
\label{fig:WSe2}
\vspace{-0.3cm}
\end{figure}

While we have chosen MoTe$_2$/EuO with its predicted huge magnetic proximity effects, it is important to explore if the main trends will persist in other TMD-based heterostructures having weaker proximity effects. Motivated by recent optical measurements~\cite{Zhao2017:NN}, we repeat our analysis of the absorption spectra for WSe$_2$/EuS. In this system, observed valley splittings of 2.5 meV for out-of-plane $B=1$ T greatly exceed the Zeeman effect of 0.1-0.2 meV/T in TMDs on nonmagnetic substrates. Consistent with our analysis, it can be attributed to magnetic proximity effects induced by the out-of-plane ${\bm M}$ in EuS. To describe WSe$_2$, we choose standard parameters~\cite{Kormanyos2015:2DM,footote:WSe2Parameters} and recall that for ${\bm M}$={\bf 0} there is a reversed CB ordering between MoTe$_2$ and WSe$_2$ [Fig~\ref{fig:Scheme}(a)]. EuS, a magnetic insulator extensively used to demonstrate robust spin-dependent effects, including the first demonstration of solid-state spin filtering~\cite{Zutic2004:RMP,Esaki1967:PRL,Moodera1988:PRL}, has significantly smaller exchange splittings than EuO, chosen in Fig.~\ref{fig:WSe2} to be $J_\mathrm{c}=12.5$ meV and $J_\mathrm{v}=5$ meV. We again neglect the Rashba SOC based on the studies of MoTe$_2$/EuO, where even at a large value, $\lambda_R=50$ meV, arising due to the structural inversion asymmetry, all the key features are preserved and just slightly blueshifted as compared to the absorption spectra in Fig.~\ref{fig:Comparison}~\cite{Note:SM}.

Focusing on the A peaks for $\sigma_+$ and $\sigma_-$, Figs.~\ref{fig:WSe2}(a) and~(b) exhibit a lifting of the valley degeneracy by an out-of-plane ${\bm M}$ with the peaks shifted oppositely for both polarizations. This behavior, experimentally observed also in Ref.~\onlinecite{Zhao2017:NN}, is similar to Fig.~\ref{fig:Comparison}, but with smaller splittings of $\sim$ 14 meV for the A and B excitonic peaks. Likewise, the dark excitons below the A peak become bright due to an in-plane ${\bm M}$, albeit with much less spectral weight lost by the A peak to the formerly dark excitons. This smaller transfer of spectral weight reflects the much smaller values of $J_\mathrm{c(v)}$ than for EuO. Figure~\ref{fig:WSe2}(c) shows the peak positions of the bright and dark excitons with the magnetization reversal from $\hat{\bm{z}}$ to $-\hat{\bm{z}}$. Similar to Fig.~\ref{fig:Rotation} and to experimental observations in Ref.~\onlinecite{Zhao2017:NN}, the relative ordering of bright, as well as dark, excitonic peaks for both $\sigma_+$ and $\sigma_-$ is reversed when ${\bm M}$ is rotated from $\hat{\bm{z}}$ to $-\hat{\bm{z}}$ (in our setup, light is propagating along $-\hat{\bm{z}}$).

Despite these similarities, there are also striking differences between the evolution of excitons in the two heterostructures. Unlike Fig.~\ref{fig:Rotation}, the bright excitons do not become dark as ${\bm M}$ is reversed in Fig.~\ref{fig:WSe2}(c)~\cite{Note:SM}. Instead, the oscillator strength of the formerly dark excitons has its maximum for in-plane ${\bm M}$ and then decays to zero as ${\bm M}$ is rotated to $-\hat{\bm{z}}$ [see the symbol sizes qualitatively representing the exciton oscillator strengths in Fig.~\ref{fig:WSe2}(c)]. This is a consequence of $J_\mathrm{c(v)}$ being dominated by $\lambda_\mathrm{c(v)}$ and hence the relative spin ordering in both valleys is unchanged compared to the case of $\bm{M}=\bm{0}$. The smaller values of $J_\mathrm{c(v)}$ in Fig.~\ref{fig:WSe2} compared to Figs.~\ref{fig:Comparison} and~\ref{fig:Rotation} also result in the shifts of the exciton peaks with $\phi$ which are noticeably smaller in Fig.~\ref{fig:WSe2}(c) than in Fig.~\ref{fig:Rotation}. Nevertheless, our calculations for WSe$_2$/EuS show that, even in TMD-based magnetic heterostructures with weaker ferromagnetic exchange, an in-plane ${\bm M}$ is expected to result in pronounced signatures of the dark excitons.

So far, magnetic proximity effects in TMDs employing ferromagnetic insulators and semiconductors were measured at cryogenic temperatures. However, this is not a fundamental limitation: Common ferromagnetic metals could enable room temperature proximity effects, while the metal/ML TMD hybridization can be prevented by inserting a thin insulating layer. A similar approach for Co- or graphene-based heterostructures that predicted a gate-controlled sign change in the proximity-induced spin polarization~\cite{Lazic2016:PRB} was recently confirmed experimentally at 300 K~\cite{Guram2017:P}, suggesting important opportunities to study unexplored phenomena in TMDs. Unlike $B$ fields of $\sim$ 30 T~\cite{Molas2017:2DM,Zhang2017:NN} that exceed typical experimental capabilities, the removal of valley degeneracy using magnetic substrates is not complicated by orbital effects and yet could enable even larger valley splittings~\cite{Qi2015:PRB}. Magnetic proximity offers another way to control and study many-body interactions in the time-reversed valleys of ML TMDs. For example, by competing with the influence of the intrinsic spin-orbit coupling, it would change the energy of shortwave plasmons~\cite{Dery2016:PRB} put forth as an explanation for the low-energy dynamic band observed in W-based electron-doped TMDs~\cite{vanTuan2017:P,Jones2013:NN}. After the completion of our work, the discovery of 2D van der Waals ferromagnets~\cite{Gong2017:N,Huang2017:N} suggests additional opportunities to probe converting excitons in TMDs.

\begin{acknowledgments}
We thank K. Belashchenko, J. Cerne, H. Dery, A. Petrou, and H. Zeng for valuable discussions. This work was mainly supported by the U.S. DOE, Office of Science BES, under Award No. DE-SC0004890, the German Science Foundation (DFG) via Grants No. SCHA 1899/2-1, SFB 689 and SFB 1170 ``ToCoTronics'', and by the ENB Graduate School on Topological Insulators. Computational work was supported by the UB Center for Computational Research and the Unity Through Knowledge Fund, Contract No. 22/15.
\end{acknowledgments}

\section{Bethe-Salpeter Equation}\label{Sec:BSE}

In this work, we employ the procedure described in Refs.~\onlinecite{Rohlfing2000:PRB,Scharf2016:PRB} to compute neutral excitons. If intervalley coupling is not taken into account, the excitons can be calculated for each valley separately. Then, neutral excitons with momentum $q_\mathrm{exc}=0$ can be described by their quantum number $S$ and their valley $\tau$ and calculated from the Bethe-Salpeter equation (BSE),
\begin{equation}\label{BSE}
[\Omega^{\tau}_{S}-\epsilon^\tau_c(\bm{k})+\epsilon^\tau_v(\bm{k})] \mathcal{A}^{S\tau}_{vc\bm{k}} =\sum\limits_{v'c'\bm{k}'}\mathcal{K}^{\tau}_{vc\bm{k},v'c'\bm{k}'}\mathcal{A}^{S\tau}_{v'c'\bm{k}'},
\end{equation}
given as Eq.~(3) in the main text~\cite{footnote:BSE}. Here, the band index $n=c(v)$ denotes one of the two conduction (valence) bands, $\epsilon^\tau_n(\bm{k})$ the corresponding single-particle/quasiparticle eigenenergies in valley $\tau$ with momentum $\bm{k}$, and $\Omega^{\tau}_{S}$ the energy of the exciton state $\ket{\Psi^{\tau}_{S}}$. We have used the ansatz $\ket{\Psi^{\tau}_{S}}=\sum_{vc\bm{k}}\mathcal{A}^{S\tau}_{vc\bm{k}}\hat{c}^\dagger_{\tau c\bm{k}}\hat{c}_{\tau v\bm{k}}\ket{\mathrm{GS}}$ for the exciton state with the coefficients $\mathcal{A}^{S\tau}_{vc\bm{k}}$, the creation (annihilation) operator of an electron in a conduction band $c$ (valence band $v$) $\hat{c}^\dagger_{\tau c\bm{k}}$ ($\hat{c}_{\tau v\bm{k}}$) in valley $\tau$, and the ground state $\ket{\mathrm{GS}}$ with fully occupied valence bands and unoccupied conduction bands.

Equation~(\ref{BSE}) describes an eigenvalue problem for the eigenvalue $\Omega^{\tau}_{S}$ and the eigenvector $\mathcal{A}^{S\tau}_{vc\bm{k}}$. The interaction kernel~\cite{Rohlfing2000:PRB}, consists of the direct and exchange terms,
\begin{equation}\label{Kernel}
\mathcal{K}^{\tau}_{vc\bm{k},v'c'\bm{k}'}=\mathcal{K}^{\mathrm{d},\tau}_{vc\bm{k},v'c'\bm{k}'}+\mathcal{K}^{\mathrm{x},\tau}_{vc\bm{k},v'c'\bm{k}'},
\end{equation}
\begin{equation}
\label{KernelDirect}
\begin{aligned}
\mathcal{K}^{\mathrm{d},\tau}_{vc\bm{k},v'c'\bm{k}'}=-\int\d^2r\d^2r'&W\left(\bm{r}-\bm{r}'\right)\left\{\left[\psi^\tau_{c\bm{k}}(\bm{r})\right]^\dagger\psi^\tau_{c'\bm{k}'}(\bm{r})\right\}\\
&\times\left\{\left[\psi^\tau_{v'\bm{k}'}(\bm{r}')\right]^\dagger\psi^\tau_{v\bm{k}}(\bm{r}')\right\},
\end{aligned}
\end{equation}
\begin{equation}\label{KernelExchange}
\begin{aligned}
\mathcal{K}^{\mathrm{x},\tau}_{vc\bm{k},v'c'\bm{k}'}=\int\d^2r\d^2r'&V\left(\bm{r}-\bm{r}'\right)\left\{\left[\psi^\tau_{c\bm{k}}(\bm{r})\right]^\dagger\psi^\tau_{v\bm{k}}(\bm{r})\right\}\\
&\times\left\{\left[\psi^\tau_{v'\bm{k}'}(\bm{r}')\right]^\dagger\psi^\tau_{c'\bm{k}'}(\bm{r}')\right\},
\end{aligned}
\end{equation}
if only intravalley Coulomb interactions are taken into account. Here, $\psi^\tau_{n\bm{k}}(\bm{r})$ denote the wave functions of the single-particle states with energies $\epsilon^\tau_n(\bm{k})$ (see above), $V(\bm{r})$ the bare Coulomb potential determined from the dielectric environment, and $W(\bm{r})$ the screened Coulomb potential, all in real space.

The single-particle eigenstates $\psi^\tau_{n\bm{k}}(\bm{r})=\exp(\i\bm{k}\cdot\bm{r})\eta^{\tau}_{n\bm{k}}/\sqrt{A}$ (with unit area $A$) are determined from $H_\mathrm{tot}\eta^{\tau}_{n\bm{k}}=\epsilon^\tau_n(\bm{k})\eta^{\tau}_{n\bm{k}}$, where the single-particle Hamiltonian ${H}_\mathrm{tot}={H}_0+{H}_\mathrm{ex}+{H}_\mathrm{R}$ is defined in the main text. Inserting $\psi^\tau_{n\bm{k}}(\bm{r})$ into Eqs.~(\ref{KernelDirect}) and~(\ref{KernelExchange}), we obtain
\begin{equation}
\label{KernelDirectEffectiveModel}
\mathcal{K}^{\mathrm{d},\tau}_{vc\bm{k},v'c'\bm{k}'}=-\frac{W\left(\bm{k}-\bm{k}'\right)f_{cc'}\left(\bm{k},\bm{k}'\right)f_{v'v}\left(\bm{k}',\bm{k}\right)}{A},
\end{equation}
\begin{equation}
\label{KernelExchangeEffectiveModel}
\mathcal{K}^{\mathrm{x},\tau}_{vc\bm{k},v'c'\bm{k}'}=-\frac{V\left(\bm{k}-\bm{k}'\right)f_{cv}\left(\bm{k},\bm{k}\right)f_{v'c'}\left(\bm{k}',\bm{k}'\right)}{A},
\end{equation}
where the form factors $f^{\tau}_{nn'}\left(\bm{k},\bm{k}'\right)=\left[\eta^{\tau}_{n\bm{k}}\right]^\dagger\eta^{\tau}_{n'\bm{k}'}$ are calculated from the single-particle states and $W(\bm{k})$ and $V(\bm{k})$ are the Fourier transforms of $W(\bm{r})$ and $V(\bm{r})$. Due to orthogonality of the eigenspinors $\eta^{\tau}_{n\bm{k}}$, $f^{\tau}_{cv}\left(\bm{k},\bm{k}\right)=0$ and $\mathcal{K}^{\mathrm{x},\tau}_{vc\bm{k},v'c'\bm{k}'}$ vanishes. Hence, we are left with $\mathcal{K}^{\tau}_{vc\bm{k},v'c'\bm{k}'}=\mathcal{K}^{\mathrm{d},\tau}_{vc\bm{k},v'c'\bm{k}'}$ in our model. In addition to the single-particle states, $W(\bm{k})$ is needed to compute $\mathcal{K}^{\tau}_{vc\bm{k},v'c'\bm{k}'}$, with expressions for the Coulomb potentials $W(\bm{k})$ and $V(\bm{k})$ discussed in Sec.~\ref{Sec:Potential} below.

\begin{figure}[t]
\vspace{0.3cm}
\includegraphics*[width=8.6cm]{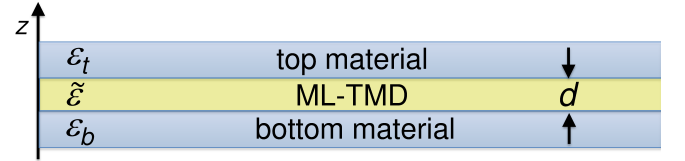}
\caption{(Color online) Geometry of the setup: A monolayer transition-metal dichalcogenide (ML TMD) of thickness $d$ and dielectric constant $\tilde{\varepsilon}$ is embedded in two other materials with dielectric constants $\varepsilon_t$ and $\varepsilon_b$.}\label{fig:Geometry}
\end{figure}

\section{Bare Coulomb Potential and Geometric Corrections}\label{Sec:Potential}

If the conduction bands are completely empty and all valence states are occupied, the screened potential and the bare Coulomb potential, that is, the potential unscreened by free charge carriers, coincide, $W(\bm{q})=V(\bm{q})$. The bare Coulomb potential is determined only from the dielectric environment (see Fig.~\ref{fig:Geometry}) and can be obtained from the Poisson equation. Then, the bare Coulomb interaction between two electrons in the $xy$-plane ($z=z'=0$) can be calculated as~\cite{Zhang2014:PRB,Scharf2016:arxiv}
\begin{widetext}
\begin{equation}\label{Coulomb_potential_exact}
\begin{aligned}
V(\bm{q})=\frac{2\pi e^2}{q}\frac{(\tilde{\varepsilon}^2-\varepsilon_t\varepsilon_b)+(\tilde{\varepsilon}^2+\varepsilon_t\varepsilon_b)\cosh(qd)+\tilde{\varepsilon}(\varepsilon_t+\varepsilon_b)\sinh(qd)}{\tilde{\varepsilon}\left[(\tilde{\varepsilon}^2+\varepsilon_t\varepsilon_b)\sinh(qd)+\tilde{\varepsilon}(\varepsilon_t+\varepsilon_b)\cosh(qd)\right]}.
\end{aligned}
\end{equation}
\end{widetext}
For thin layers $qd\ll1$, $1/V(\bm{q})$ can be expanded in powers of $qd$, which yields
\begin{equation}\label{Coulomb_potential_approx}
V(\bm{q})\approx
\frac{2\pi e^2}{\varepsilon q+r_0q^2},
 \end{equation}
where
$\varepsilon=(\varepsilon_t+\varepsilon_b)/2$
is the average dielectric constant of the bottom and top materials surrounding the monolayer transition-metal dichalcogenide (ML TMD), and
\begin{equation}\label{Coulomb_potential_approx_r0}
r_0=\frac{\tilde{\varepsilon}d}{2}\left(1-\frac{\varepsilon_t^2+\varepsilon_b^2}{2\tilde{\varepsilon}^2}\right)
\end{equation}
can be interpreted as the polarizability of the monolayer. In the limit of $\varepsilon_{t/b}\ll\tilde{\varepsilon}$, $r_0=\tilde{\varepsilon}d/2$ and we recover the result derived in Refs.~\onlinecite{Keldysh1979:JETP,Cudazzo2011:PRB}.

The interaction given by Eqs.~(\ref{Coulomb_potential_approx})-(\ref{Coulomb_potential_approx_r0}) has proven to be highly successful in capturing the excitonic properties of ML TMDs~\cite{Berkelbach2013:PRB} and is also used in our calculations. With $\varepsilon_t=1$ for air, $\varepsilon_b=23.9$ for EuO and $\varepsilon_b=11.1$ for EuS, we obtain average $\varepsilon=12.45$ for air/MoTe$_2$/EuO and $\varepsilon=6.05$ for air/WSe$_2$/EuS from and model the geometric correction by $r_0=6.3$ nm for air/MoTe$_2$/EuO and by $r_0=4.0$ nm for air/WSe$_2$/EuS.

\begin{figure}[ht]
\centering
\vspace{0.3cm}
\includegraphics*[width=8.6cm]{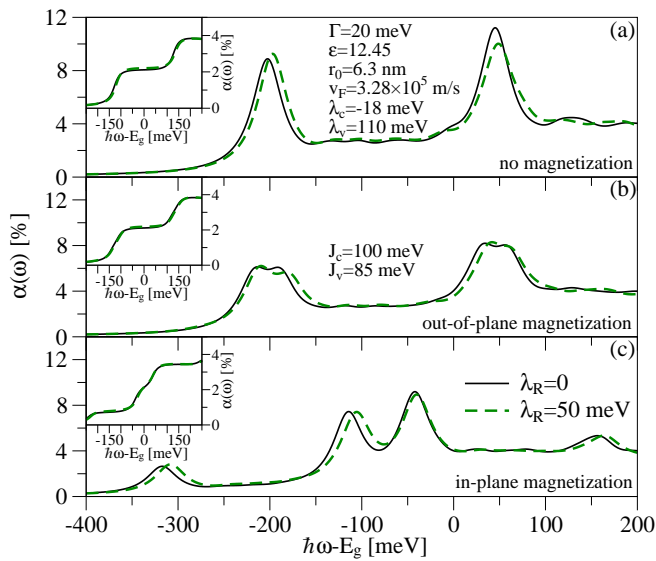}
\caption{Absorption spectra for linear polarization of ML MoTe$_2$ on EuO without (solid lines) and with (dashed lines) Rashba SOC for (a) no proximity-induced exchange splitting, (b) out-of-plane and (c) in-plane exchange splitting. The insets show the respective single-particle absorptions computed without excitonic effects. $E_\mathrm{g}=1.4$ eV.}
\label{fig:ComparisonRashba}
\end{figure}

\section{Effect of Rashba Spin-orbit Coupling}\label{Sec:Rashba}

In the main text, we have presented results for $\lambda_\mathrm{R}=0$. The reason for this is twofold: First, even for a large value such as $\lambda_R=50$ meV, all key features discussed in the absence of Rashba spin-orbit coupling (SOC) are preserved. The second reason is that it allows for a simpler description of the effect the magnetic substrate has on the absorption spectrum of a ML TMD and that it is possible to provide simple analytical estimates and formulas in this case.

Figure~\ref{fig:ComparisonRashba} compares the linear absorption of ML MoTe$_2$ on EuO if Rashba SOC is neglected and if it is taken into account. In the absence of a magnetization breaking time-reversal symmetry there are only two optically active excitons, even for $\lambda_\mathrm{R}\neq0$ [Fig.~\ref{fig:ComparisonRashba}(a)]. Both for an out-of-plane [Fig.~\ref{fig:ComparisonRashba}(b)] and in-plane magnetization [Fig.~\ref{fig:ComparisonRashba}(c)], Rashba SOC does also not qualitatively change the behavior of the absorption. The main quantitative change due to Rashba SOC is that the absorption spectra are just slightly blue-shifted as compared to the absorption spectra in Fig.~2 in the main text.

\section{Computational Details}\label{Sec:Computational}

In our model, where intervalley coupling is neglected and excitons can be calculated for each valley separately, Eq.~(\ref{BSE}) [Eq.~(3) in the main text] contains two conduction and two valence bands for a given valley $\tau$. In order to diagonalize Eq.~(\ref{BSE}) [Eq.~(3) in the main text], we use a coarse uniform $N\times N$ $k$-grid with a spacing of $\Delta k=2\pi/(Na_0)$ and $a_0=3.52$ {\AA} for MoTe$_2$ and $a_0=3.29$ {\AA} for WSe$_2$ in each direction as well as an upper energy cutoff $E_\mathrm{cu}$. The Coulomb matrix elements $W\left(\bm{k}-\bm{k}'\right)$, however, are not evaluated at the grid points of the $N\times N$ $k$-grid, but instead are averaged over a square centered around the coarse grid point $\bm{k}-\bm{k}'$ with side widths $\Delta k$ on a fine $N_\mathrm{int}\times N_\mathrm{int}$ grid [with a corresponding spacing of $\Delta k_\mathrm{int}=\Delta k/N_\mathrm{int}=2\pi/(NN_\mathrm{int}a_0)$]. To ensure convergence in our numerical calculations, we have used $N=60$, $N_\mathrm{int}=100$ and energy cutoffs of 500 meV above the band gap, $E_\mathrm{cu}=E_\mathrm{g}/2+500$ meV, for air/MoTe$_2$/EuO and 750 meV above the band gap, $E_\mathrm{cu}=E_\mathrm{g}/2+0.5$ eV, for air/WSe$_2$/EuS. This procedure ensures that our numerical calculations converge reasonably fast and we find that the $1s$ binding energy changes by less than $1\%$ when going from $N=60$ to $N=100$.

\section{Absorption profile of WSe2}\label{Sec:WSe2}

\begin{figure}[ht]
\centering
\includegraphics*[width=8.6cm]{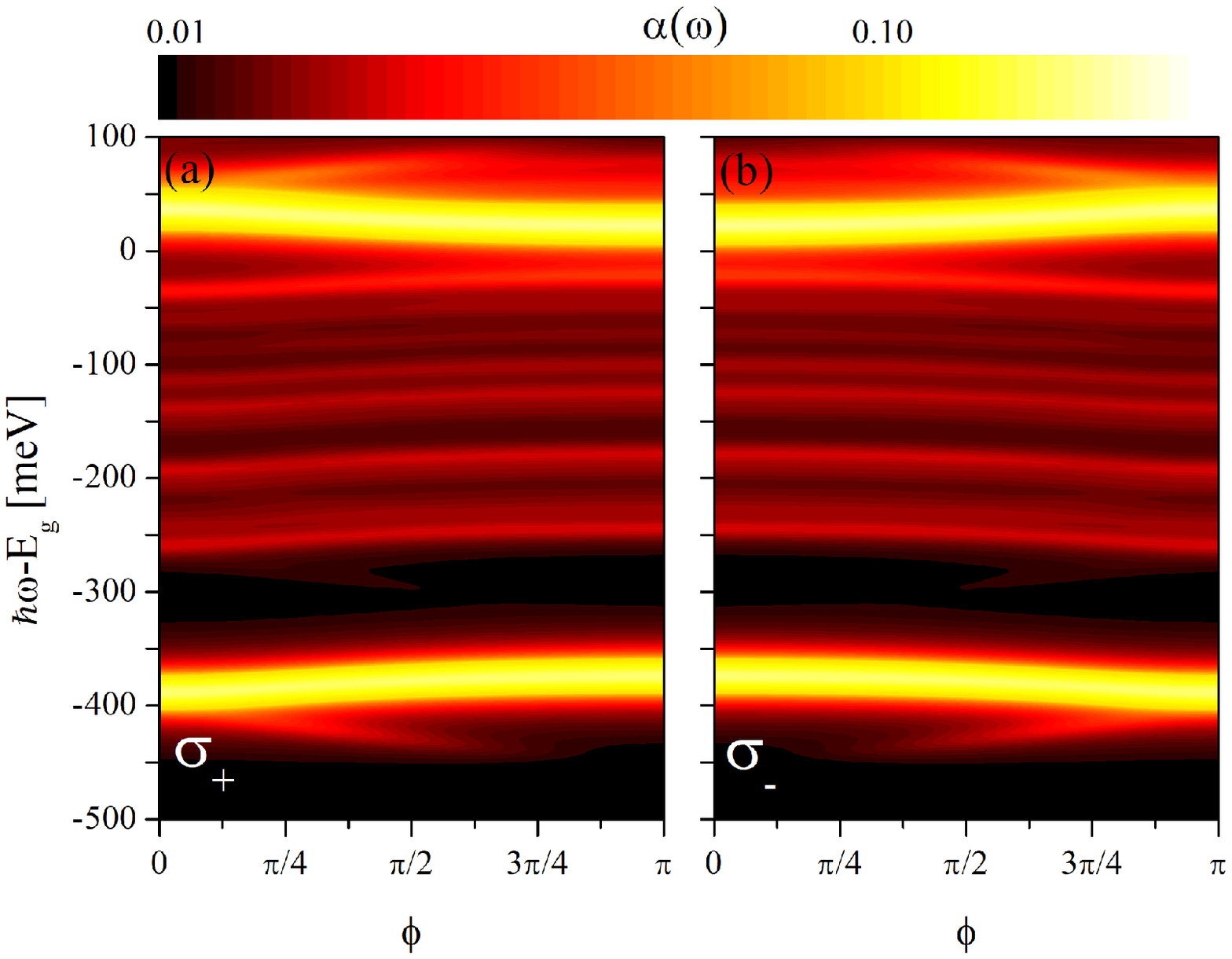}
\vspace{-0.5cm}
\caption{Evolution of the absorption of ML WSe$_2$ on EuS as ${\bm M}$ is rotated from out of plane ($\phi=0$) to in plane ($\phi=\pi/2$) and out of plane, but with reversed ${\bm M}$ ($\phi=\pi$) for (a) $\sigma_+$ and (b) $\sigma_-$.The parameters are the same as in Fig.~4 of the main text with a broadening of $\Gamma=10$ meV.}
\label{fig:WSe2Rotation}
\vspace{-0.3cm}
\end{figure}

As noted in the main text, the absorption of WSe$_2$/EuS exhibits a qualitatively different behavior than the one predicted for MoTe$_2$/EuO due to the significantly smaller exchange splittings in WSe$_2$/EuS. In Fig.~4 of the main text, we have focused on the A peaks of the $\sigma_+$ and $\sigma_-$ absorption for ML WSe$_2$ on EuS. To fully illustrate the differences between the two systems, Fig.~\ref{fig:WSe2Rotation} shows the absorption of ML WSe$_2$ on EuS as ${\bm M}$ is rotated from out of plane ($\phi=0$) to in plane ($\phi=\pi/2$) and out of plane, but with reversed ${\bm M}$ ($\phi=\pi$). In contrast to MoTe$_2$/EuO (compare to Fig.~3 of the main text), there is no conversion from dark to bright excitons and vice versa. Instead, the dark excitons become bright as ${\bm M}$ is rotated in plane and turn dark again as ${\bm M}$ is rotated out of plane, but with reversed ${\bm M}$.

\section{Magnetization effects on the Bethe-Salpeter Equation}\label{Sec:MBSE}

The effect of the magnetization ${\bm M}$ is taken into account by the single-particle Hamiltonian~(2) in the main text. Hence, the ${\bm M}$-dependence in our approach enters the BSE~(\ref{BSE}) on the right-hand side (RHS) via the single-particle energies $\epsilon^\tau_{c/v}(\bm{k})$ and on the left-hand side (LHS) via the form factors $f^{\tau}_{nn'}\left(\bm{k},\bm{k}'\right)$ in the kernel [see Eq.~(\ref{KernelDirectEffectiveModel})].

\begin{figure}[t]
\centering
\subfloat[Binding energies $E_\mathrm{b}(\phi)$]{\includegraphics*[width=7cm]{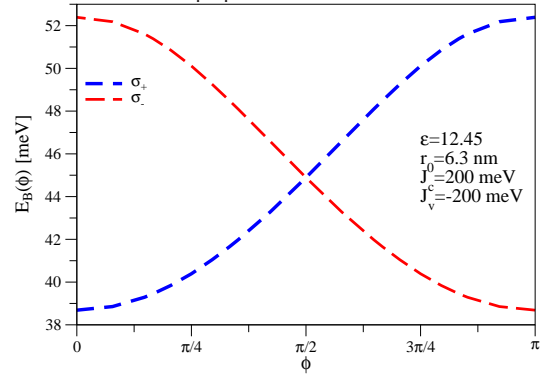}}\\
\subfloat[Relative changes of $E_\mathrm{b}(\phi)$]{\includegraphics*[width=7cm]{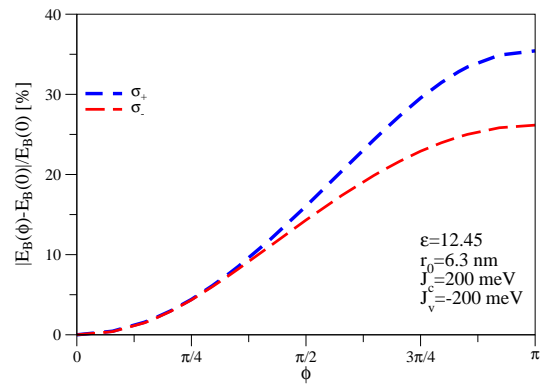}}
\caption{Evolution of (a) the binding energies, $E_\mathrm{b}(\phi)=[\epsilon^\tau_c(\phi)-\epsilon^\tau_v(\phi)]-\Omega^{\tau}_{S}(\phi)$ and (b) their relative change $|E_\mathrm{b}(\phi)-E_\mathrm{b}(0)|/E_\mathrm{b}(0)$ for $v_\mathrm{F}=3.28\times10^5$ m/s, $\lambda_\mathrm{c}=-18$ meV, $\lambda_\mathrm{c}=110$ meV, $\varepsilon=12.45$, $r_0=6.3$ nm, $J_{\mathrm{c}}=-J_{\mathrm{v}}=200$ meV.}\label{fig:AlternateMoTe2}
\end{figure}

As a consequence of these two effects, the exciton peak is, in general, not just following the evolution of the single-particle excitation spectrum, but is subject to additional corrections. These corrections can be traced by following the binding energy $E_\mathrm{b}(\phi)=[\epsilon^\tau_c(\phi)-\epsilon^\tau_v(\phi)]-\Omega^{\tau}_{S}(\phi)$. We have noted in the main text that, for the used parameters of ML TMDs on magnetic substrates, these corrections are small and the full BSE results can be well described by a rigid shift of the excitonic peaks given by the single-particle spin splitting (see Fig. 3 of the main text). However, the single-particle description gives neither position nor the shape of the absorption peaks.

In order to illustrate that this `single-particle' rigid shift agreement is not universal, Figs.~\ref{fig:AlternateMoTe2}(a) and~(b) show the binding energies $E_\mathrm{b}(\phi)$ and their relative change $|E_\mathrm{b}(\phi)-E_\mathrm{b}(0)|/E_\mathrm{b}(0)$ as ${\bm M}$ is rotated for a different set of parameters. Here, we have chosen $v_\mathrm{F}=3.28\times10^5$ m/s, $\lambda_\mathrm{c}=-18$ meV, $\lambda_\mathrm{c}=110$ meV, $\varepsilon=12.45$, and $r_0=6.3$ nm, similar to MoTe$_2$/EuO, but with $J_{\mathrm{c}}=-J_{\mathrm{v}}=200$ meV and $E_\mathrm{g}=1.0$ eV. The binding energy changes by as much as 30\% and exhibits a more pronounced dependence on ${\bm M}$ than for the typical parameters shown in the main text of the paper.

Table~\ref{tab:parameters} contains $E_\mathrm{b}(\phi)$ for out-of-plane and in-plane ${\bm M}$ and different combinations of $J_{\mathrm{c}}$, $J_{\mathrm{v}}$ and $E_\mathrm{g}$. It illustrates that, in principle, there are parameter regimes where rigidly shifting the single-particle excitation energies by a constant, ${\bm M}$-independent binding energy is not a good description of the exciton peaks.

\begin{widetext}

\begin{table}
\begin{center}
\begin{tabular}{|c||c|c|c|}
\hline
 parameters & $E_\mathrm{b}(0)$ & $E_\mathrm{b}(\pi/2)$ & $E_\mathrm{b}(\pi)$\\
\hline\hline
$E_\mathrm{g}=1.0$ eV, $J_c=200$ meV, $J_v=-200$ meV & 38.6 meV & 44.9 meV & 52.4 meV\\
\hline
$E_\mathrm{g}=0.7$ eV, $J_c=200$ meV, $J_v=-200$ meV & 22.2 meV & 27.2 meV & 36.1 meV\\
\hline\hline
$E_\mathrm{g}=1.0$ eV, $J_c=200$ meV, $J_v=-100$ meV & 44.4 meV & 50.2 meV & 63.3 meV\\
\hline\hline
$E_\mathrm{g}=1.0$ eV, $J_c=100$ meV, $J_v=-200$ meV & 44.3 meV & 49.8 meV & 57.0 meV\\
\hline\hline
$E_\mathrm{g}=1.0$ eV, $J_c=200$ meV, $J_v=-50$ meV & 47.0 meV & 53.0 meV & 63.9 meV\\
\hline\hline
$E_\mathrm{g}=1.0$ eV, $J_c=50$ meV, $J_v=-200$ meV & 47.0 meV & 51.9 meV & 59.2 meV\\
\hline
$E_\mathrm{g}=1.0$ eV, $J_c=-200$ meV, $J_v=100$ meV & 63.3 meV & 50.2 meV & 44.4 meV\\
\hline\hline
\end{tabular}
\end{center}
\caption{Comparison between the binding energies of the lowest exciton for $\sigma_+$ for out-of-plane and in-plane magnetizations for ML MoTe$_2$ on EuO ($v_\mathrm{F}=3.28\times10^5$ m/s, $\lambda_\mathrm{c}=-18$ meV, $\lambda_\mathrm{c}=110$ meV, $\varepsilon=12.45$, $r_0=6.3$ nm) if the gap parameter $E_\mathrm{g}$ and/or the exchange splittings $J_{\mathrm{c/v}}$ are changed. The binding energies for $\sigma_-$ are the same as those for $\sigma_+$ for in-plane magnetization and reversed for out-of-plane magnetization.}\label{tab:parameters}
\end{table}

\end{widetext}


\begin{thebibliography}{99}
\bibitem{footnote:BSE} We do not take into account intervalley scattering processes in this work.
\end{thebibliography}
\end{document}